\documentclass[titlepage,12pt]{article}
\usepackage{amssymb}
\usepackage{amsfonts}
\usepackage{amsmath}
\begin{document}

{\bf \Large How Far Should the Principle \\ \\ of Relativity Go ?} \\ \\

{\bf Elem\'{e}r E Rosinger} \\
Department of Mathematics \\
and Applied Mathematics \\
University of Pretoria \\
Pretoria \\
0002 South Africa \\
eerosinger@hotmail.com \\ \\

{\bf Abstract} \\

The Principle of Relativity has so far been understood as the {\it covariance} of laws of
Physics with respect to a general class of reference frame transformations. That relativity,
however, has only been expressed with the help of {\it one single} type of mathematical
entities, namely, the scalars given by the usual continuum of the field $\mathbb{R}$ of real
numbers, or by the usual mathematical structures built upon $\mathbb{R}$, such as the scalars
given by the complex numbers $\mathbb{C}$, or the vectors in finite dimensional Euclidean
spaces $\mathbb{R}^n$, infinite dimensional Hilbert spaces, etc. \\
This paper argues for {\it progressing deeper and wider} in furthering the Principle of
Relativity, not by mere covariance with respect to reference frames, but by studying the
possible covariance with respect to a {\it large variety} of algebras of scalars which extend
significantly $\mathbb{R}$ or $\mathbb{C}$, variety of scalars in terms of which various
theories of Physics can equally well be formulated. \\
First directions in this regard can be found naturally in the simple Mathematics of Special
Relativity, the Bell Inequalities in Quantum Mechanics, or in the considerable amount of
elementary Mathematics in finite dimensional vector spaces which occurs in Quantum
Computation. \\
The large classes of algebras of scalars suggested, which contain $\mathbb{R}$ and
$\mathbb{C}$ as particular cases, have the important feature of typically {\it no longer}
being Archimedean, see Appendix, a feature which can prove to be valuable when dealing with
the so called "infinities" in Physics. \\
The paper has a Comment on the so called "end of time". \\ \\

{\bf 1. The Status-Quo, and Going Beyond ...} \\

As argued in the sequel, Theoretical Physics has for long by now been confined to the following
status-quo, without however being aware of that confinement, and instead, taking it for granted
as the only obvious and natural way :

\begin{itemize}

\item The scalars used are based on the set $\mathbb{R}$ of real numbers, upon which the
complex numbers $\mathbb{C}$, as well as various finite dimensional Euclidean, or for that
matter, infinite dimensional Hilbert spaces are constructed.

\item As it happens, $\mathbb{R}$ is algebraically a {\it field}, see Appendix, that is,
addition, subtraction, multiplication and division can be made without restriction, except for
division with zero.

\item As it also happens, $\mathbb{R}$ is algebraically {\it Archimedean}, which means among
others that it {\it cannot} conveniently operate with infinity or distinguish between variants
of infinity, some more infinite than other ones. This leads to the well known theoretical
difficulties called "infinities in physics".

\end{itemize}

There have been attempts to go beyond such a confinement by considering the use of scalars
other than $\mathbb{R}$ or $\mathbb{C}$. Such attempts, however, were found not particularly
encouraging in view, among others, of the following :

\begin{itemize}

\item The alternative scalars were required to be again {\it fields}, just as is the case with
$\mathbb{R}$ and $\mathbb{C}$.

\item The use of {\it non-Archimedean} scalars has been seen as too difficult, thus they tended
to be avoided.

\end{itemize}

As it turns out - and a fact which appears not to be particularly familiar with theoretical
physicists - the previous two points are simply just about {\it incompatible} with one another.
Indeed, according to a classical theorem of Pontrjagin, [22], the only fields which  are not
discrete are $\mathbb{R}$, $\mathbb{C}$ and the Hamiltonian quaternions $\mathbb{H}$, the
latter being non-commutative, however, Archimedean as well. \\

Therefore, in case one is interested in a genuine and rich enough class of new possible
scalars, one has to {\it abandon} the requirement that such scalars constitute a field. \\

Now, as it happens, such a requirement is in fact not at all hard for theoretical physicists,
or for that matter, engineers and economists, or others as well, since they have for long been
familiar with the use of {\it matrices}. And certainly, the square matrices of any given order
$n\geq 2$ do {\it no longer} constitute a field, since there are plenty of nonzero matrices one
cannot divide with, if their determinants vanish. \\
However, with square matrices of a given order $n \geq 2$, one can still do unrestricted
addition, subtraction and multiplication, as well as division, except for those with zero
determinant. \\

In this way, square matrices constitute an algebraic structure which is called {\it algebra},
see Appendix, a structure which is but slightly more general, hence only marginally less rich
or convenient, than that of a field. \\

And thus if we are ready to give up looking for fields - a choice severely restricted by the
Pontrjagin theorem - then we need not go farther but to the very next milder algebraic
structure, namely, of {\it algebras}, a structure physicists, engineers, economists, and others
as well, have in fact been familiar with for quite a while by now ... \\

And once we are ready for that step, a {\it very large} class of easy to construct and use
{\it algebras} turn out to be available, as shown in the sequel, algebras which are {\it
commutative} as well. \\
And as one of the unexpected additional advantages, these algebras turn out to be {\it
non-Archimedean}. \\

And why their non-Archimedean algebraic structure is actually an advantage, when theoretical
physicists tend to think otherwise ? \\

Well, as argued in the sequel, and so far not much familiar with theoretical physicists, the
fact is that, contrary to accepted views, we simply do {\it not} have the freedom of choice
between Archimedean and non-Archimedean algebraic structures. Indeed, the former prove to be
but minuscule subsets of the latter. Thus by choosing to work exclusively with the former, we
confine ourselves without ever becoming aware of that fact ... \\
Ever, except perhaps when, like for instance in Theoretical Physics, we keep hitting time and
again upon the difficulties brought about by the so called "infinities in physics" ... \\

And then, what may be the way forward ? \\

Well, the existence of the mentioned very large class of algebras - called {\it reduced power
algebras} - offers not only the possibility to redo much of Physics in their terms, but also
the following one, so far not considered, see [14], which could bring with it a significant
further {\it deepening and widening} of the Principle of Relativity itself :

\begin{itemize}

\item To explore the extent to which theories of Physics are, or on the contrary, are not
independent of the respective scalars they use in their mathematical formulation.

\end{itemize}

{~} \\

{\bf 2. Main Moments in the Evolution of the Principle of \\
        \hspace*{0.45cm} Relativity} \\

The Principle of Relativity in Physics has so far undergone the following {\it three stages} :

\begin{itemize}

\item Aristotelian and pre-Galilean,

\item Galillean - Newtonian and of Special Relativity,

\item General Relativity.

\end{itemize}

Here, following several recent papers, [8-10,14,15,17,18], a {\it fourth stage} will be
suggested, namely one which takes the Principle of Relativity

\begin{itemize}

\item from reference frame relativity to the relativity of mathematical models involved in the
theories of Physics.

\end{itemize}

In the first, Aristotelian and pre-Galilean stage there was simply {\it no} Principle of
Relativity, but precisely its very opposite. Indeed, it was considered that Planet Earth was
immobile, and of course, at the very center of the Universe. And as recalled in the sequel,
Aristotle believed to have a perfectly valid proof of it. That view was closely related to
Aristotle's assumption that, when formulated in modern terms, velocity - and not acceleration
- was proportional to force in motion. Since in ancient Greek times no experiments of any more
substantive nature were made in this regard, and since they could not avail themselves of
Calculus, no one realized the immediately resulting contradiction. Namely, such an assumption
would mean that the law of motion would be given by a {\it first order} differential equation
in terms of position. Thus contrary to obvious and elementary facts of common experience, one
would only be able to impose one single initial condition on motion, be that, for instance, an
initial position, or an initial velocity. Certainly, one could not impose two independent
initial conditions, say, both an initial position and an initial velocity, as is the case with
Newton's law of motion, his second law in fact, which is given by a {\it second order}
differential equation in terms of position. \\

It was Galileo, with his by now classical argument about moving with constant velocity in the
belly of a boat on a still lake which, as far as known, introduced for the first time the idea
of the Principle of Relativity for the motion of objects within Classical Mechanics. This
Galilean Principle of Relativity was fully taken over by Newton, and expressed in his mentioned
law of motion, according to which acceleration - and not velocity - in motion is proportional
with force. This law, as mentioned, is given by a {\it second order} differential equation in
terms of position, thus it is perfectly compatible with the commonly known possibility of
being able to impose no less than two independent initial conditions. \\

Einstein's Special Relativity took over the Galilean - Newtonian Principle of Relativity, and
extended it to Electro-Magnetism as well. And in fact, as mentioned later, as far as Special
Relativity is concerned, this relativity axiom alone is sufficient, since it does imply a
finite and constant speed for light in vacuum. \\

The third stage in the evolution of the Principle of Relativity in Physics occurred with
Einstein's General Relativity. And this stage introduced a {\it massive deepening and
widening} of that principle. Indeed, this time that principle was no longer limited to inertial
reference frames, as in its original Galilean-Newtonian and Special Relativity versions. \\

Moreover, a further essential difference between the above stages two and three is the
following. \\

Both Classical Mechanics and Special Relativity are {\it background dependent}. In other
words, within these theories the space-time is given a priori and once and for ever, as a four
dimensional Euclidean algebraic structure of vector space, being thus totally independent of
the physical processes which take place in it. In this way, that specific background is in
fact {\it forced} upon those theories of Physics, which therefore do not have any freedom, but
to depend on it. \\
In sharp contradistinction to that, General Relativity is {\it background independent}, since
it creates its own space-time background as given by the respective solution of the Einstein
equations, once a distribution of masses is specified. \\

And then, in line with the mentioned three above stages undergone by the Principle of
Relativity, and with an aim to explore its possible further extensions, one may rather
naturally ask :

\begin{itemize}

\item How far and deep {\it background independence} may actually go ? Is there still some
given background upon which even General Relativity happens to depend ?

\end{itemize}

Unrelated to such questions, a rather natural other question has for a while by now started to
insinuate itself into the physical thinking, as seen for instance in the literature cited in
[8,9,14,15,17,18]. Namely, there have been questions raised with respect to

\begin{quote}

{\it what scalars should be used in Physics} ?

\end{quote}

beyond and above the usual real and complex numbers. \\

However, we can look at that question within a far deeper and wider context, namely, of the
possible further extensions of he Principle of Relativity, and thus, a further diminishing of
background dependence of theories of Physics. In other words, we can try to strive {\it not}
only for a reference frame relativity, as achieved so far and in a considerable measure in
General Relativity, but also for a relativity with respect to the very {\it mathematical
models} used in the theories of Physics. And needless to say, within this context, the {\it
scalars} used in theories of Physics - scalars which are involved in the construction of so
many other entities in theories of Physics, among them space-time for instance - are some of
the obvious first elements of such mathematical models which come naturally to attention. \\

As seen in the sequel, there exists indeed a considerable variety of scalars which can be used
in theories of Physics, namely, the so called {\it reduced power algebras}, which turn out to
be rather natural extensions of the usual real and complex numbers. \\
And thus there exists an effective opportunity to explore the extent to which theories of
Physics are indeed, or on the contrary, are not independent of the respective scalars they use
in their mathematical formulation. A first step in this regard was outlined and pursued to an
extent in [14], as somewhat earlier suggested in [8,9]. \\

As it happens, there are important theories of Physics where a good deal of the arguments only
employ relatively simple Mathematics. Such is the case, for instance, with Special Relativity,
the Bell Inequalities in Quantum Mechanics, or for that matter, in the considerable amount of
rather elementary Mathematics in finite dimensional vector spaces which occurs in Quantum
Computation. An example of a nontrivial classical theory of physical interest is that of Chaos,
where the one dimensional case is rather well understood, and specifically, the role played by
the two Feigenbaum constants alpha and delta. Such a theory, which again, contains a good deal
of quite elementary Mathematics, could be subjected to a reformulation in terms of reduced
power algebras, with a corresponding study of what remains valid, and what becomes
different. \\

Here one should immediately note the considerable {\it relevance} both a positive, and
alternatively, a negative answer to such a mathematical model independence may offer. \\
Indeed, in the case of a positive answer, one would obtain a significant extension and
deepening of the Principle of Relativity. And based on that, one could further pursue, possibly
by other means than the scalar algebras suggested here or in [8,9,14,15,17,18], the study of
the extension and deepening of the Principle of Relativity. \\
On the hand, a negative answer would immediately raise rather fundamental questions, not least
among them about the status of those laws of Physics which fail to be independent with respect
to the scalar algebras suggested here. \\

Finally, let us mention that as a byproduct of considering algebras of reduced powers, one is
naturally led to scalars which are {\it non-Archimedean}, see Appendix for the algebraic
notions and properties used in the sequel. And such non-Archimedean scalars differ considerably
from both the real and complex numbers in their far more easy, convenient, and above all,
sophisticated ways of dealing with "infinities". \\
In this regard, one simply is made aware of the fact that it has been but an historical
accident that we ended up with the Archimedean scalars of the usual real and complex numbers,
thus with considerable difficulties in dealing with "infinities". \\

And what one discovers in the process is that Archimedean algebraic structures, and in
particular scalars, such as the usual real and complex numbers, are but most particular {\it
subsets} of by no means less natural non-Archimedean algebraic structures. \\
This is precisely the reason why "infinities" - so troubling in theories of Physics - do
inevitably appear in Archimedean algebraic structures, namely, due to the simplistic approach
such algebraic structures exhibit, when compared with the non-Archimedean ones. \\
In this regard, one may recall how primitive human tribes would have a counting system that
would only know about the following distinctions :

\begin{quote}

"one", "two", "three", and "many" ...

\end{quote}

And as it turns out, a somewhat similar situation occurs with Archimedean algebraic structures
which, the moment one reaches somewhat farther towards large quantities, are only able to
record that in one and only one way, namely, as a "blow up" which puts an instant end to all
algebraic operations, since one has reached a so called "infinity" ... \\
On the other hand, and in sharp contradistinction, non-Archimedean algebraic structures, such
as for instance the reduced power algebras in the sequel, are sophisticated enough in order
to be able to incorporate without any difficulty within their algebraic operations such
situations which are simply "no go" realms for Archimedean algebraic structures. \\

Consequently, one finds that in fact, one does {\it not} have the freedom of choice between
Archimedean and non-Archimedean algebraic structures, see [15,17,18] in this regard, since the
choice of the former inevitably confines one to a particular and highly inconvenient situation,
even if it has been the one we happened to choose historically a long time ago, and have
limited ourselves to it ever since ... \\ \\

{\bf 3. Was That an Equivocation with Relativity ... ?} \\

As it happens, Relativity Theory, both in its Special and General versions, and introduced by
Einstein in the early 1900s, is considered along the not much later originated Quantum Theory,
as being the two truly revolutionary theories of modern Physics. And yet, as far as Relativity
is concerned, its fundamental idea, the very idea which is reflected in its name, has not
always been seen with enough clarity. Indeed, Einstein himself, when formulating his theory of
Special Relativity, [2-5], set at its foundations two axioms, the first of which is about what
is in fact an extension of the classical Galilean Principle of Relativity, incorporating this
time Electro-Magnetism as well, while the second is about the velocity of light in vacuum.
Here we reproduce in the translation the respective section which is in the preamble to his
famous 1905 paper [2] : \\

\begin{quote}

"Examples of this sort ... suggest that the phenomena of electrodynamics as well as of
mechanics possess no properties corresponding to the idea of absolute rest. They suggest rather
that, as has already been shown to the first order of small quantities, the same laws of
electrodynamics and optics will be valid for all frames of reference for which the equations of
mechanics hold good. We will raise this conjecture(the purport of which will hereafter be
called the "Principle of Relativity") to the status of a postulate, and also introduce another
postulate, which is only apparently irreconcilable with the former, namely, that light is
always propagated in empty space with a definite velocity $c$ which is independent of the state
of motion of the emitting body."

\end{quote}

And here we already have a first instance of what may be seen as a sort of equivocation with
respect to the idea of Relativity, since soon after, [6,19], it was shown that Einstein's
second axiom is in fact a {\it consequence} of the first one. \\

A second equivocation, and in fact, a somewhat more manifest and significant one, occurred
during the next decade, when Einstein tried to include in Relativity gravitation as well, and
he did so based on the axiom of equivalence between gravitational and inertial mass. In this
process, however, Einstein had several well known unsuccessful attempts, until in the late
1915 he found the formulation of General Relativity which has been accepted ever since. \\

And why can these two episodes be seen as equivocating attempts at incorporating the Principle
of Relativity into Physics ? \\

Well, in the case of the first one, this follows from the very fact that the two respective
axioms are actually not independent, and with the second one considered by Einstein {\it not}
being recognized by him as a mere consequence of the first one. As is well known from
Einstein's biography, the issue of the velocity of light had preoccupied him for most of the
previous decade, thus it is not surprising to have taken a position among his two axioms, even
if it turned out to be dependent on the first one. Certainly, Einstein was above all a
physicists, and as such, he was no doubt fascinated with the phenomenon of light, thus giving
it a position which, as an independent axiom, it proved simply {\it not} to have. \\

As it turned out, this deep attachment to physical intuition had a yet greater influence on
the way Einstein was to reach General Relativity. Indeed, instead of simply searching for
those laws of Physics which are {\it covariant} not merely with respect to Lorentz, but with
respect to as general as possible reference frame transformations, Einstein placed a
considerable importance on the Principle of Equivalence. The effect was the well known series
of unsuccessful attempts, prior to late 1915, in reaching the correct form of the General
Relativity. \\

This somewhat equivocating approach to the Principle of Relativity is further highlighted by
the fact that in various later publications aimed at a larger readership, [3,4], Einstein
placed an obvious stress on that principle, considering it to be the fundamental one. \\

In our own days, however, there is an increased clarity, [20], about the fact that the
Principle of Relativity is indeed fundamental, and it should be seen as a {\it general
covariance} property of the laws of Physics. And among others, this means the {\it background
independent} nature of such laws, and in general, of theories of Physics. \\ \\

{\bf 4. But, How Deep and Wide Does the {\it Background} Go ?} \\

For a better understanding of the Principle of Relativity in what it may be its further and
yet more full relevance in Physics, it may seem appropriate to start questioning our usual
assumptions involved in its present formulation. And as it turns out, some of such assumptions
are so usual in fact, as to be accepted by us through a less than conscious mental, if not
even emotional, reflex. And needless to say, our long human record in regard of having such
kind of assumptions, and on top of it, of dealing with them in such a reflex manner, is
certainly clear and well established, even if not often enough up front in our awareness. \\
After all, and as one of the many blatant examples, we can recall how many even among the most
learned and considered to be wise sages did, for ages and up until Copernicus, do nothing else
but take absolutely for granted the assumption that Planet Earth was immobile at the center of
the Universe ? \\
Aristotle himself firmly believed to have a most simple and incontrovertible proof of it :
when one drops a stone from a tall tower, it always falls at the foot of that tower ... \\
So much for our long historical record with respect to understanding the Principle of
Relativity ... \\

So then, let us try to see which may be the backgrounds presently involved, be it consciously
or less so, in the formulation of the Principle of Relativity.

\begin{itemize}

\item
One of the deepest backgrounds, no doubt, is that of physical intuitions, a source which often
is so much valued by physicists as to lead to its much preferred top priority - if not in fact
nearly exclusive - use, and then, like with the mentioned two cases involving Einstein, to
certain less than best possible processes in thinking. \\
Amusingly, such a much preferred just about exclusively exclusive reliance on physical
intuition firmly rejects taking even one single page from the Nobel Laureate physicist Eugene
Wigner's celebrate 1960 argument about the ... unreasonable effectiveness of mathematics in
natural sciences ... \\
This background of physical intuition, however, is at present clearly outside of an approach
sufficiently accessible to science, as far as its more relevant ways of functioning are
concerned. \\
Nevertheless, the essential role of this background in setting up the theories of Physics is
all too obvious, as shown by the history of science which records ever new insights as well as
a better understanding of older ones, as times goes on.

\item
A less deep, yet still fundamental background is that of the {\it logic} used. And this is
already at a level which, ever since Aristotle, and even more so in our times, has been the
subject of considerable scientific study. In this regard, what is used in theories of Physics
is, so far, exclusively the usual binary valued logic, with the rule of the excluded middle,
and without allowing circular arguments. \\
In this regard, however, there are already two remarkable {\it novelties}. Namely, rigorous
mathematical theories have been developed, [1], in which circular arguments play an essential
role. Also, rigorous mathematical theories in which contradictions are allowed, [7,21], have
been a subject of study. \\
In view of that, it may perhaps come the time in the not too distant future for certain
theories of Physics to embrace logical structures such as already studied and applied in
[1,7,21].

\item
However, before venturing into realms of nonclassical logic which still seem so strange to
many, there is another, and nearer to us, fundamental level which, so far, has seldom been
subjected to enquiry. Namely, and as mentioned, see also [8,9,14,15,17,18] and the references
cited there, it is about the issue of :

\begin{itemize}

\item What scalars should we use in theories of Physics ?

\end{itemize}

\end{itemize}

Indeed, in their decreasing order of depth as background to theories of Physics, physical
intuition and logic are naturally followed by the {\it scalars} used in such theories. And
if due to various possible reason, among them those mentioned above, we may not yet be ready
to question or reconsider the first two, as they happen to be used at present, then perhaps
even more so it may be high time to do such a reconsideration with the third one, namely, with
the scalars presently used in theories of Physics. \\

And if we are to pursue consistently the Principle of Relativity, then we should realize the
following :

\begin{itemize}

\item The physical intuition of physicists is by its nature an extraordinarily rich and
creative source. Therefore, if not relied upon too exclusively or narrowly, it can hardly
conflict with the {\it background free} nature of the Principle of Relativity.

\item The presently used logic in the theories of Physics, on the other hand, by its very
uniqueness, by its very exclusivity, does inevitably {\it nail down} a specific background,
either we like it or not, either we are aware of it or not. And the question is wide open, and
in fact, hardly ever considered, whether holding to such a fixed background, no matter how
natural it may seem, may in fact conflict with the background free nature of the Principle of
Relativity.

\item Reconsidering the scalars used in theories of Physics remains, therefore, at present the
only workable alternative to conforming more deeply to the Principle of Relativity, and doing
so beyond the present view of it as only reduced to the covariance of laws of Physics with
respect to large classes of reference frame transformations.

\item And as it happens, such a reconsideration of the scalars used in theories of Physics is
significantly facilitated by the easily accessible and usable {\it abundance} of a large
variety of algebras available for that purpose, as indicated in the sequel, see also
[8,9,14,15,17,18,23-38,45-48], as well as the particular case in [49], and the literature
cited there.

\item Finally, once theories of Physics are reformulated in such algebras, [14], a consistent
pursuit of the Principle of Relativity requires the study of the corresponding {\it extended}
concept of covariance, namely, to what extent laws of Physics do, or on the contrary, do not
depend on the specific scalars used in the respective theories.

\end{itemize}

Contrary to what many physicists may tend to believe, the issue is not about whether Physics,
or for that matter, Mathematics is the primary scientific venture. Therefore, the issue is not
in any way of a mere partisan nature. \\
As it happens, however, few studies, if any, have been conducted about the more fundamental
possible interactions between Physics and Mathematics. \\
On the part of mathematicians, and even more so of many prominent ones, the role of Physics in
inspiring and developing outstanding new Mathematics has been well known and made fruitful use
of for a long time by now. \\
The problem, therefore, seems to be more on the side of the physicists. And too many of them
tend to see Mathematics as a sort of unpleasant exercise which, unfortunately in their
perception, cannot always be avoided. \\

Needless to say, there have been even in recent times noted exceptions on both sides involved.
For instance, the fundamentally important mathematical field of Category Theory, introduced in
the 1940s, happened not to be embraced by many prominent mathematicians. Among them was the
well known French group under the collective name of N Bourbaki, whose eventual demise in the
1980s is considered by many to have been caused among others by their systematic refusal to
redo Mathematics in terms of Category Theory, and thus go beyond its present foundation on Set
Theory. \\
What is even more amusing is that mathematicians specializing in Category Theory were not those
who recently started a most massive extension of that theory, by introducing what is called
N-Categories. Indeed, certain more abstract minded physicists from Quantum Field Theory
happened to be the originators of that latest development. \\
So much for trying to draw clear and sharp lines in such an issue as the more fundamental
possible interactions between Physics and Mathematics ... \\

A rather unique and most impressive moment happened with the mentioned paper of Eugene Wigner,
entitled "The Unreasonable Effectiveness of Mathematics in the Natural Sciences",
Communications in Pure and Applied Mathematics, Vol. 13, No. 1, February 1960, which raised a
more fundamental issue in the pursuit of modern science, one not limited only to the
interaction between Physics and Mathematics. \\
As it happened, however, there was hardly any notable debate following that paper ... \\
Quite everybody in Natural Sciences seemed to be far too busy with trying to pursue their own
interests, and do so with their own specific means, of which Mathematics would, when on occasion
unavoidable, be seen rather as a necessary but unpleasant detour ... \\

As for Wigner's mentioned paper, itself does not seem to go deep enough, beyond the
exemplification of that unreasonable effectiveness of Mathematics, and towards the possible
reasons for it. \\

And indeed, which may be such possible reasons ? \\

Well, one way to see the whole issue is perhaps as follows. \\
First of all, we should realize that Mathematics got a wrong naming, since its essence is in
no way reflected in it. \\
And what is the essence of Mathematics ? \\
Well, quite likely that, so far, it is the only science developed by us humans which is

\begin{itemize}

\item both {\it abstract}

\item and {\it precise}.

\end{itemize}

Philosophy, for instance, is certainly abstract at its best, however, it is quite far from
being precise as well. With Physics, on the other hand, the situation is the other way round,
since it is rather precise, but clearly not abstract enough, not enough even to include, say,
Chemistry, let alone, Biology, and so on. \\

And then, it may be that it is the {\it abstract} nature of Mathematics, in conjunction, of
course, with its {\it precise} character which makes it so widely effective in Natural
Sciences. After all, Philosophy is even more widely relevant, due to its abstract nature. Yet
lacking precision, it is not of much effective help on more detailed levels in Natural
Sciences. \\

And in case such a view of Mathematics may indeed have enough merit, then the interest
physicists should exhibit in it would come precisely from the deeper and more general patterns
Mathematics can access, patterns which often go far beyond more usually accessible insights,
and patterns which may nowadays be not only useful, but also quite necessary to physicists, if
not in fact, of fundamental importance, given the increasingly counter-intuitive nature of much
of modern Physics, starting with Relativity, and including of course the rather mysterious
realms of Quanta ... \\

In this way, it is a rather open question, and quite likely to remain as such for a long time
to come, how deep and wide may the background go, with respect among others to the Principle
of Relativity ? \\ \\

{\bf 5. A Long Ongoing Reflex Ancient Egyptian and Archimedean \\
        \hspace*{0.45cm} Choice ...} \\

Ever since Euclid's geometry, as later algebrized by Descartes, there has been one and only
one choice of scalars used in Physics, namely, as given by the usual continuum of the field
$\mathbb{R}$ of real numbers. Indeed, the complex numbers $\mathbb{C}$, the finite dimensional
Euclidean spaces $\mathbb{R}^n$, or even infinite dimensional Hilbert spaces, etc., are all
built upon the real numbers $\mathbb{R}$. \\

No less that five fundamental features of the real numbers $\mathbb{R}$ play an important role,
one that has for long by now been taken for granted. Namely, $\mathbb{R}$ is a :

\begin{itemize}

\item {\it field},

\item {\it linearly}, or {\it totally} ordered,

\item {\it Archimedean}, see Appendix,

\item {\it complete} topologically, in other words, a {\it continuum}, and

\item the {\it only one} with the above four properties.

\end{itemize}

Such a list of credentials seems indeed more than enough to confer upon $\mathbb{R}$ a
position as {\it the} undisputed exclusive mathematical model to be used in theories of
Physics. \\
Not to mention that on top of it, and according to a well known result obtained in the 1930s
by Pontrjagin, [22], the only fields which are not totally disconnected are $\mathbb{R},~
\mathbb{C}$ and the Hamiltonian quaternions $\mathbb{H}$, the latter being noncommutative. \\

Nevertheless, such an impressive list of credential should rather be left to explain the
reasons why, historically, we happened to come across the real and complex numbers, than to
end up by confining us for evermore to their exclusive use in the theories of Physics. \\
After all, as noted, a more consistent pursuit of the background free nature of the Principle
of Relativity is not supposed to acquiesce in the exclusive use of no matter which only one
logic. Therefore, it is even less supposed to do so to the exclusive use of one single, no
matter how naturally looking, set of scalars. \\

There are, however, a number of additional reasons why we nevertheless ended up with the real
numbers $\mathbb{R}$ playing such a uniquely fundamental role in theories of Physics. One of
them comes from the fact that $\mathbb{R}$ is a {\it field}, that is, the operations of
addition, subtraction, multiplication and division can be effectuated freely, except for
division with zero. \\
Also, the multiplication in $\mathbb{R}$ is conveniently {\it commutative}. \\

Here however, we can note that physicists, and even engineers, have for long by now been
accustomed to dealing with {\it matrices}. And square matrices in general, that is, of a given
order $n \geq 2$, are {\it not} a field, since there are plenty of such nonzero matrices which
have their determinant zero, thus they do not have an inverse, and then one cannot divide with
them. Furthermore, the multiplication of such matrices is in general {\it not} commutative. \\

Such algebraic structures in which addition, subtraction and multiplication can be done
without restrictions, while division cannot always be done with nonzero elements are called
{\it algebras}. Fields are, therefore, particular cases of algebras. \\
Some of the algebras, like for instance those of matrices of a given order $n \geq 2$, are
noncommutative. However, there are plenty of commutative algebras as well, and a large class
of them will be presented in the sequel. \\ \\

{\bf 6. There Are Plenty of Scalar Algebras Easy to Construct \\
        \hspace*{0.5cm} and Use ...} \\

It appears, therefore, that one of the reasons why physicists have been so much limiting
themselves to the use of the real numbers $\mathbb{R}$ and of the structures built upon them
is that the real numbers constitute a {\it field}, and thus, one can do unrestricted divisions
with such numbers, the only exception being division with zero. \\

And yet as is well known, the real numbers $\mathbb{R}$ are {\it not} the only field
available. Therefore, if so much tempted by the advantages of working with scalars in a field,
then why not choose scalars in some other field for formulating the theories of Physics ? \\

The answer, although not quite clearly stated, or for that matter, consciously enough known by
physicists is that the other fields available do happen to involve certain {\it difficulties}.
Indeed, let us recall here the two classical results already mentioned in this regard :

\begin{itemize}

\item The theorem according to which $\mathbb{R}$ is the only field which is totally ordered,
topologically complete, thus, it is a continuum, and it is also Archimedean.

\item The theorem of Pontrjagin, according to which the real numbers $\mathbb{R}$, the complex
numbers $\mathbb{C}$ and the quaternions $\mathbb{H}$ are the only fields which are not
totally disconnected.

\end{itemize}

It follows, therefore, that all other possible fields must inevitably {\it fail} to be a
continuum and/or Archimedean. And indeed, the field $^*\mathbb{R}$ of nonstandard real numbers,
for instance, fails to be Archimedean, while the various p-adic fields fail to be a
continuum. \\
Added to the above difficulties come also the technically involved manner such fields are
constructed, a manner which makes their use rather cumbersome, when compared with the use of
the usual real numbers in $\mathbb{R}$. \\

And then, precisely here comes in the possibility to set aside the use of scalars in a field,
and instead, use scalars in a {\it large class of algebras} presented in the sequel. \\
And the advantages in doing so will be as follows :

\begin{itemize}

\item Obtaining an easy to construct and use large setup within which we can consider the
further extension and deepening of the Principle of Relativity, and do so this time not only
with respect to reference frame transformations or the usual background independence of the
type encountered in General Relativity, but also within the significantly more general concept
of background independence with respect to the mathematical models which give the scalars used
in the theories of Physics.

\item Doing away with the long ongoing and difficult issue of "infinities in physics", a thus
as well with the need for the rather ill-founded variety of methods called "renormalization".

\item Becoming aware of the fact that we do {\it not} have the freedom of choice to avoid
dealing with scalars which belong to non-Archimedean algebras.

\end{itemize}

{~} \\

{\bf 7. The Simple Algebraic Construction of the \\
        \hspace*{0.5cm} Large Class of {\it Reduced Power Algebras}} \\

In order to diminish the possible difficulties for physicists, we start by pointing out that,
fortunately, our fundamental building block for reduced power algebras is still the familiar
set $\mathbb{R}$ of the usual scalars given by the real numbers. \\

{\bf Power Algebras} \\

Added to that, there are only {\it two} new ingredients. The first one is any {\it infinite}
set $\Lambda$ of indices $\lambda \in \Lambda$. The second one will come not much later. \\

The {\it first step} in constructing the reduced power algebras is to go from the usual real
number scalars in $\mathbb{R}$, which by the way, algebraically constitute a {\it field}, to
the {\it vastly larger algebra} that is no longer a field, namely the {\it power algebra} \\

(7.1)~~~ $ \mathbb{R}^\Lambda $ \\

and which, as known in Set theory, can naturally be identified with the set of all sequences
of real numbers with indices in $\Lambda$, namely \\

(7.2)~~~ $ \xi = ( \xi_\lambda )_{\,\lambda \,\in\, \Lambda},~~~ \mbox{where}~~~
                                                            \xi_\lambda \in \mathbb{R} $ \\

Alternatively, and equally naturally, $\mathbb{R}^\Lambda$ can be seen as the set of all real
valued functions defined on $\Lambda$, that is \\

(7.3)~~~ $ \xi : \Lambda \longrightarrow \mathbb{R},~~~ \mbox{where}~~~ \Lambda \ni \lambda
                                \longmapsto \xi ( \lambda) = \xi_\lambda \in \mathbb{R} $ \\

Now the way $\mathbb{R}^\Lambda$ is an {\it algebra} is as follows. \\

Given two sequences $\xi = ( \xi_\lambda )_{\,\lambda \,\in\, \Lambda},~ \eta =
( \eta_\lambda )_{\,\lambda \,\in\, \Lambda} \in \mathbb{R}^\Lambda$, their {\it addition} is
defined term-wise, that is, by \\

(7.4)~~~ $ \xi + \eta = ( \xi_\lambda + \eta_\lambda)_{\,\lambda \,\in\, \Lambda} \in
                                                                  \mathbb{R}^\Lambda $ \\

and similarly, one defines term-wise their {\it multiplication}, namely \\

(7.5)~~~ $ \xi \,.\, \eta = ( \xi_\lambda \,.\, \eta_\lambda)_{\,\lambda \,\in\, \Lambda} \in
                                                                  \mathbb{R}^\Lambda $ \\

Finally, the multiplication of sequences in $\mathbb{R}^\Lambda$ with real number scalars from
$\mathbb{R}$ is also defined term-wise. Thus given a real number scalar $\alpha \in
\mathbb{R}$ and a sequence $\xi = ( \xi_\lambda )_{\,\lambda \,\in\, \Lambda} \in
\mathbb{R}^\Lambda$, we define \\

(7.6)~~~ $ \alpha \, \xi = ( \alpha \, \xi_\lambda )_{\,\lambda \,\in\, \Lambda} \in
                                                            \mathbb{R}^\Lambda $ \\

It should be pointed out that all of the above definitions of algebraic operations in (7.4) -
(7.6) are but standard. Moreover, in terms of the representation of $\mathbb{R}^\Lambda$ in
(7.3) as a set of real valued functions on $\Lambda$, these operations are precisely the usual
ones with functions. \\

Consequently, there should not be any unease with the algebra $\mathbb{R}^\Lambda$. \\

And why is $\mathbb{R}^\Lambda$ only an algebra, and {\it not} a field as well ? \\

Simple, namely, there are many elements, that is, sequences in $\mathbb{R}^\Lambda$ which are
{\it not} identically zero, yet one cannot divide with them. This fact is easier to follow if
we use the function representation (7.3) for the elements of $\mathbb{R}^\Lambda$. Indeed, as
is well known, given any function \\

(7.7)~~~ $ \xi : \Lambda \longrightarrow \mathbb{R} $ \\

it is {\it not} possible to divide with that function $\xi$, unless it {\it never} vanishes,
that is, unless it satisfies the condition \\

(7.8)~~~ $ \xi ( \lambda ) \neq 0,~~~ \mbox{for all}~~ \lambda \in \Lambda $ \\

And obviously, since $\Lambda$ contains at least two different elements, being assumed to be
in fact an infinite set, there are many functions (7.7) in $\mathbb{R}^\Lambda$ which vanish
on some nonvoid part of $\Lambda$, yet do not vanish on another nonvoid part of it. Thus such
functions are not identically zero, yet they fail  to satisfy (7.8), and then, one can no
longer divide with them. \\

For clarity, let us give a simple example, when $\Lambda = \mathbb{N}$. In this case, the
sequence \\

$~~~~~~ \xi = ( 1, 0, 1, 0, 1, 0, \ldots ) \in \mathbb{R}^\Lambda = \mathbb{R}^\mathbb{N} $ \\

is evidently not identically zero, yet one cannot divide by $\xi$, and in particular, $1 /
\xi$ is not well defined, since $\xi$ happens to contain terms which are zero, thus fails to
satisfy (7.8). \\

In view of its form, see (7.1), which is standard notation in Set Theory, the algebra
$\mathbb{R}^\Lambda$ can be seen as a {\it power algebra}, obtained from the real numbers
$\mathbb{R}$ by exponentiation with the infinite index set $\Lambda$. \\

What is important to note is that $\mathbb{R}^\Lambda$ is an algebra {\it extension} of the
usual real numbers in $\mathbb{R}$. Indeed, to every real number $x \in \mathbb{R}$, let us
associate the following sequence of real numbers, sequence indexed by indices in $\Lambda$,
namely \\

(7.9)~~~ $ u_x = ( v_\lambda )_{\,\lambda \,\in\, \Lambda} \in \mathbb{R}^\Lambda,~~~
                 \mbox{where}~~ v_\lambda = x,~~~ \mbox{for}~~ \lambda \in \Lambda $ \\

thus $u_x$ is the {\it constant} sequence whose terms are each the same, namely, $x$. Then
one obtains the following {\it algebra embedding}, that is, an injective algebra
homomorphism \\

(7.10)~~~ $ \mathbb{R} \ni x \longmapsto u_x \in \mathbb{R}^\Lambda $ \\

In this extension of $\mathbb{R}$, however, the power algebra $\mathbb{R}^\Lambda$ is {\it
immensely larger} than  the real numbers $\mathbb{R}$, due to the fact that $\Lambda$ is an
infinite set. In this regard, let us note that in case $\Lambda$ would be a finite set with $n
\geq 2$ elements, the corresponding power algebra $\mathbb{R}^\Lambda$ would be - as a vector
space, and without the multiplication operation in (7.5) - the $n$-dimensional Euclidean space
$\mathbb{R}^n$, which itself is already considerably larger than $\mathbb{R}$, for $n$ large
enough. \\
In this way, the power algebras $\mathbb{R}^\Lambda$, with $\Lambda$ an infinite set, can be
seen, when considered to be vector spaces, as infinite dimensional extensions of usual
Euclidean spaces. \\

As for the algebra embedding (7.10), we note that in case $\Lambda$ would have only two
elements, thus the power algebra $\mathbb{R}^\Lambda$ would be as a vector the two dimensional
Euclidean space $\mathbb{R}^2$, then the set $\{~ u_x ~~|~~ x \in \mathbb{R} ~\}$, which is the
range of the embedding (7.1), would be precisely the {\it diagonal} subset of
$\mathbb{R}^2$. \\

Furthermore, it is easy to see that in the general case of (7.10), the set \\

(7.11)~~~ $ {\cal U}\,_\Lambda = \{~ u_x ~~|~~ x \in \mathbb{R} ~\} \subset \mathbb{R}^\Lambda $ \\

is a {\it subalgebra} of $\mathbb{R}^\Lambda$, and as an algebra, it is {\it isomorphic} with
$\mathbb{R}$. \\

{\bf Reducing the Power Algebras} \\

And now comes the {\it second and last step} in the construction of the reduced power
algebras. \\

For that, we take as a {\it second} ingredient any {\it proper ideal} ${\cal I}$ in the power
algebra $\mathbb{R}^\Lambda$, that is \\

(7.12)~~~ $ {\cal I} \subsetneqq \mathbb{R}^\Lambda $ \\

and construct the {\it quotient algebra}, see the respective standard method in the
Appendix \\

(7.13)~~~ $ \mathbb{A} = \mathbb{R}^\Lambda / {\cal I} $ \\

which quite appropriately is called a {\it reduced power algebra}. Indeed, its construction
can now be summarized as follows : \\

In addition to the usual set $\mathbb{R}$ of real numbers, it contains two ingredients, namely

\begin{itemize}

\item an {\it infinite} index set $\Lambda$ which is used to construct from $\mathbb{R}$ the
immensely larger power algebra $\mathbb{R}^\Lambda$, and

\item a {\it proper ideal} ${\cal I}$ in the power algebra $\mathbb{R}^\Lambda$ with which to
{\it reduce} by a standard quotient construction the algebra $\mathbb{R}^\Lambda$, and thus
obtain the aimed at {\it reduced power algebra} $\mathbb{A} = \mathbb{R}^\Lambda / {\cal I}$.

\end{itemize}

{\bf The Abundance in the Amount of Reduced Power Algebras} \\

The resulting abundance in the amount of reduced power algebras available will now be made
obvious, as it results from the freedom to choose the infinite index sets $\Lambda$, and  also
the proper ideals ${\cal I} \subset \mathbb{R}^\Lambda$. \\

Related to the choice of the index sets $\Lambda$ there is no need for further comments,
except to mention how the various resulting reduced power algebras may relate to one another.
This issue has been studied and presented in [9]. \\

Here we recall the way the choice of the proper ideals ${\cal I} \subset \mathbb{R}^\Lambda$
contributes to the abundance in the amount of reduced power algebras available. That issue was
also studied and presented in [9]. \\

The {\it essential} and {\it most convenient} property of the proper ideals ${\cal I} \subset
\mathbb{R}^\Lambda$ is that they can be put into a direct {\it one-to-one} correspondence with
the much simpler mathematical entities given by {\it filters} ${\cal F}$ on the index sets
$\Lambda$, see Appendix for the concept of filter. \\

This two-way correspondence happens as follows. \\

Given any proper ideal ${\cal I} \subset \mathbb{R}^\Lambda$, we associate with it the
following filter on $\Lambda$ \\

(7.14)~~~ $ {\cal F}_{\cal I} = \{~ Z ( \xi ) ~~|~~ \xi \in {\cal I} ~\} $ \\

where for $\xi \in \mathbb{R}^\Lambda$, one denotes $Z ( \xi ) = \{ \lambda \in \Lambda ~|~
\xi ( \lambda ) = 0 \}$, that is, the so called zero set of $\xi$. \\

Conversely, given any filter ${\cal F}$ on $\Lambda$, we associate with it the proper ideal \\

(7.15)~~~ $ {\cal I}_{\cal F} =
                 \{~ \xi \in \mathbb{R}^\Lambda ~~|~~ Z ( \xi ) \in {\cal F} ~\} $ \\

Let us show here how easy it is to prove (7.14) and (7.15). \\

For (7.14), we have to show, see Appendix, that ${\cal F}_{\cal I} \neq \phi$. But this
results immediately from the fact that ${\cal I} \neq \phi$. \\

Further, we have to show that $\phi \notin {\cal F}_{\cal I}$. Assume therefore that $Z ( \xi
) = \phi$, for some $\xi \in {\cal I}$. Then obviously $\xi ( \lambda ) \neq 0$, for $\lambda
\in \Lambda$, hence $1 / \xi$ is well defined, and $1 / \xi \in \mathbb{R}^\Lambda$. However,
since ${\cal I}$ is an ideal in $\mathbb{R}^\Lambda$, it follows that \\

$~~~~~~ \xi \,.\, ( 1 / \xi ) \in {\cal I} \, \mathbb{R}^\Lambda \subseteq {\cal I} $ \\

But obviously $\xi \,.\, ( 1 / \xi ) = u_1$, thus $u_1 \in {\cal I}$, which means that \\

$~~~~~~ \mathbb{R}^\Lambda \subseteq {\cal I} \, \mathbb{R}^\Lambda \subseteq {\cal I} $ \\

thus we obtain the contradiction that ${\cal I}$ is not a proper ideal in
$\mathbb{R}^\Lambda$. \\

Also, we have to show that \\

$~~~~~~ I, J \in {\cal F}_{\cal I} ~\Longrightarrow~ I \cap J \in {\cal F}_{\cal I} $ \\

Let therefore $\xi, \eta \in \mathbb{R}^\Lambda$, then obviously \\

$~~~~~~ Z ( \xi ) \cap Z ( \eta ) = Z ( \xi^2 + \eta^2 ) $ \\

while $\xi^2 + \eta^2 \in {\cal I}$, since ${\cal I}$ is an ideal in $\mathbb{R}^\Lambda$. \\

Finally, we prove that \\

$~~~~~~ I \in {\cal F}_{\cal I},~ I \subseteq J \subseteq \Lambda
                               ~\Longrightarrow~ J \in {\cal F}_{\cal I} $ \\

Indeed, let $\xi \in \mathbb{R}^\Lambda$, such that $Z ( \xi ) \subseteq J$. Let now $\eta \in
\mathbb{R}^\Lambda$ be the characteristic function of $J$ in $\Lambda$. Then $\xi \,.\, \eta
\in {\cal I}$, since ${\cal I}$ is an ideal in $\mathbb{R}^\Lambda$, hence $Z ( \xi \,.\,
\eta ) \in {\cal F}_{\cal I}$. And now we note that $Z ( \xi \,.\, \eta ) = J$, thus indeed
$J \in {\cal F}_{\cal I}$. \\ \\

{\bf Remark} \\

It is important to note that in the proof of the relation \\

$~~~~~~ Z ( \xi ) \cap Z ( \eta ) = Z ( \xi^2 + \eta^2 ) $ \\

we essentially used the fact that the respective functions $\xi, \eta : \Lambda
\longrightarrow \mathbb{R}$ have values in $\mathbb{R}$, that is, are {\it real valued}. In
other words, we used the property of {\it real numbers}, according to which, for $x, y \in
\mathbb{R}$, we have \\

$~~~~~~ x^2 + y^2 = 0 ~\Longleftrightarrow~ x = y = 0 $ \\

a property which, for instance, is {\it not} true for complex numbers.

\hfill $\Box$ \\

As for (7.15), let $\xi, \eta \in \mathbb{R}^\Lambda$, then obviously \\

$~~~~~~ Z ( \xi + \eta ) \supseteq Z ( \xi ) \cap Z ( \eta) $ \\

therefore \\

$~~~~~~ \xi, \eta \in {\cal I}_{\cal F} ~\Longrightarrow~ \xi + \eta \in {\cal I}_{\cal F} $ \\

Also it is easy to see that \\

$~~~~~~ Z ( \xi \,.\, \eta ) \supseteq Z ( \xi ) $ \\

therefore \\

$~~~~~~ \xi \in {\cal I}_{\cal F} ~\Longrightarrow~ \xi \,.\, \eta \in {\cal I}_{\cal F} $ \\

Further, for $x \in \mathbb{R}$, we obviously have \\

$~~~~~~ Z ( x \,.\, \xi ) \supseteq Z ( \xi ) $ \\

thus \\

$~~~~~~ \xi \in {\cal I}_{\cal F} ~\Longrightarrow~ x \,.\ \xi \in {\cal I}_{\cal F} $ \\

In this way ${\cal I}_{\cal F}$ is indeed an ideal in $\mathbb{R}^\Lambda$, and it only
remains to show that it is also a proper ideal, that is \\

$~~~~~~ {\cal I}_{\cal F} \varsubsetneqq \mathbb{R}^\Lambda $ \\

Assuming on the contrary that above we have equality, then obviously $u_1 \in {\cal I}_{\cal
F} $, thus $Z ( u_1 ) \in {\cal F}$. However $Z ( u_1 ) = \phi$, thus $\phi \in {\cal F}$,
which is a contradiction. \\

Now, the above constructed one-to-one correspondence between proper ideals ${\cal I}$ in
$\mathbb{R}^\Lambda$, and on the other hand, the much simpler filters ${\cal F}$ on
$\Lambda$, namely \\

(7.16)~~~ $ {\cal I} \longmapsto {\cal F}_{\cal I},~~~ {\cal F} \longmapsto
                                                         {\cal I}_{\cal F} $ \\

has the following important properties. First, this correspondence when iterated twice, it
returns ideals into the same ideals, and filters into the same filters, namely  \\

(7.17)~~~ $ {\cal I} \longmapsto {\cal F}_{\cal I} \longmapsto {\cal I}_{{\cal F}_{\cal I}} =
                {\cal I},~~~~~
              {\cal F} \longmapsto {\cal I}_{\cal F} \longmapsto {\cal F}_{{\cal I}_{\cal F}}=
                    {\cal F} $ \\

Second, it is monotonous in the following sense. Given ${\cal I},~ {\cal J}$ two ideals in
$\mathbb{R}^\Lambda$, and ${\cal F},~ {\cal G}$ two filters on $\Lambda$, then \\

(7.18)~~~ $ \begin{array}{l}
                     {\cal I} \subseteq {\cal J} ~\Longrightarrow~ {\cal F}_{\cal I}
                                       \subseteq {\cal F}_{\cal J} \\ \\
                     {\cal F} \subseteq {\cal G} ~\Longrightarrow~ {\cal I}_{\cal F}
                                        \subseteq {\cal I}_{\cal G}
            \end{array} $ \\

In view of the above it follows that the reduced power algebras in (7.13) are in fact of the
specific form \\

(7.19)~~~ $ \mathbb{A} = \mathbb{R}^\Lambda / {\cal I}_{\cal F} $ \\

where ${\cal F}$ are arbitrary filters on $\Lambda$. For convenience, we shall use the notation \\

(7.20)~~~ $  \mathbb{A}_{\cal F} = \mathbb{R}^\Lambda / {\cal I}_{\cal F} $ \\

In this way, the {\it abundance} of the reduced power algebras (7.13), (7.20) is given by the
arbitrariness of the infinite index sets $\Lambda$ and of the corresponding filters ${\cal F}$
on these index sets. \\

{\bf Natural Homomorphisms between Reduced Power Algebras} \\

Given the mentioned abundance in reduced power algebras, the question arises what possible
relationships can be established between them ? \\

Here we show {\it two natural} families of algebra homomorphisms which exists between various
reduced power algebras, depending of the infinite index sets $\Lambda$ and on the corresponding
filters ${\cal F}$ on these index sets which, according to (7.19) define these reduced power
algebras. \\

First, we fix an infinite index set $\Lambda$ and consider two filters on it, namely \\

(7.21)~~~ $ {\cal F} \subseteq {\cal G} $ \\

Then (7.18), (7.17) yield the {\it surjective algebra homomorphism} \\

(7.22)~~~ $ \mathbb{A}_{\cal F} \ni \xi + {\cal I}_{\cal F} \longmapsto
                           \xi + {\cal I}_{\cal G} \in \mathbb{A}_{\cal G} $ \\

which means that the algebra $\mathbb{A}_{\cal G}$ is {\it smaller} than the algebra
$\mathbb{A}_{\cal F}$, more precisely \\

(7.23)~~~ $ \mathbb{A}_{\cal G} ~~\mbox{and}~~ \mathbb{A}_{\cal F} /
             ( {\cal I}_{\cal G} / {\cal I}_{\cal F} ) ~~\mbox{are isomorphic algebras} $ \\

Now, let us consider the case of two infinite index sets $\Lambda \subseteq \Gamma$. Then for
every filter ${\cal F}$ on $\Gamma$ which satisfies the condition \\

(7.24)~~~ $ \Lambda \in {\cal F} $ \\

we have the {\it surjective algebra homomorphism}, see [14, pp. 14,15] \\

(7.25)~~~ $ \mathbb{A}_{\cal F} \ni \xi + {\cal I}_{\cal F} \longmapsto
                           \xi|_\Lambda + {\cal I}_{( {\cal F}|_\Lambda )} \in
                                               \mathbb{A}_{( {\cal F}|_\Lambda )} $ \\ \\

{\bf 8. Should the Background Independence Go as Deep and \\
        \hspace*{0.5cm} Wide as Independence of Reduced Power Algebras ?} \\

We are now in a position to offer a first tentative answer to the question

\begin{quote}

"But How Deep and Wide Does the Background Go ?"

\end{quote}

in the title of section 4 above. Indeed, we have seen the following features of the reduced
power algebras (7.17), namely

\begin{itemize}

\item the ease in their construction and use,

\item their abundance in terms of the infinite index sets and corresponding filters which
define them,

\item the large amount of natural algebra homomorphism among these algebras, as their defining
index sets and corresponding filters change.

\end{itemize}

Consequently, it may appear as natural to study the extent to which the laws of Physics may,
or for that matter, may not be {\it independent} of the particular reduced power algebras
(7.17), when their elements are used as {\it scalars} instead of the usual real or complex
numbers. \\

And as mentioned above, both a positive and negative answer would have its interest. \\

A few first steps in this regard were suggested recently in [14]. \\ \\

{\bf 9. Comment on the End of Time and other \\
        \hspace*{0.5cm} Background Independent Attempts} \\ \\

{\bf 9.1. Preliminaries} \\

It is argued that a background independence of theories of Physics in which all background is
totally eliminated is in fact a throw back to the assumption of a unique and universally valid
background. Consequently, as a more genuine background independence, it is suggested that
theories of Physics should simultaneously be formulated in multiple backgrounds, and then
proven to be independent of them. A rather natural suggestion for such multiple backgrounds is
given. \\ \\

{\bf 9.2. Alternatives for Implementing Background \\
          \hspace*{0.8cm} Independence} \\

The idea that fundamental theories of Physics should be {\it background independent} has been
gaining recognition, and it can be seen as one of the most important lasting legacies of
Einstein's General Relativity with respect to foundational issues in Physics, [20]. \\

Newtonian Mechanics, as much as Special Relativity are {\it not} background independent, since
they are formulated in an a priori given four dimensional Euclidean vector space,
corresponding to one time dimension and three space dimensions. Certainly, the way time and
space are seen in Newtonian Mechanics, on one hand, and Special Relativity, on the other, are
very different. \\
However, both these theories assume as a starting point the existence of the mentioned four
dimensional vector space which is isomorphic with $\mathbb{R}^4$, even if it has its specific
additional Minkowskian geometry in the case of the latter. \\

In contradistinction to such a situation, General Relativity is the first, and so far the only
fundamental and widely accepted theory of Physics which does {\it not} start with the
assumption of any a priori given and universally valid space-time background. Instead, that
theory itself is each time setting up its specific space-time background which results from
solving the Einstein equations for every given particular distribution of masses. \\

Nevertheless, such a background independence like that exhibited by General Relativity appears
to some theoretical physicists involved in Quantum Gravity, for instance, as been
insufficient. And indeed, its main feature is {\it not} the inexistence of any background at
all, but rather the inexistence of a unique and universally valid background, such as happens
in the case of Newtonian Mechanics and Special Relativity. \\

Certainly, each of these three theories operates in a {\it uniquely} given space-time
background, the difference with General Relativity being that the respective background is
{\it no} longer universally valid and given once and for all, but it is determined each time in
a unique manner, depending on the given distribution of masses. \\

The above situation being quite clear in the respective literature, the {\it problem} starts
with the ways it is attempted to be transcended in the name of a more significant background
independence. \\

Namely, typically, a {\it total}, thus {\it extreme negation} of all possible backgrounds is
suggested, the consequence being that {\it only} various  possible {\it relationships} between
physical entities are supposed to exist, and they are only allowed to do so as if within a
{\it perfectly empty} background. \\
No wonder that the corresponding inevitable massive {\it reduction} of the structures involved
to an unprecedented level of {\it barrenness} has so far had the rather unintended and
undesirable effect of not giving much chance for the elaboration of theories able to match
anywhere near the rich complexity of many of the known physical phenomena. \\
Not to mention the fact, usually missed, that the total lack of any background is, after all,
and inevitably, a background itself. And to add to it, it is a {\it unique} and {\it
universally} valid one, thus it is a throw back to no less than the situation in Newtonian
Mechanics and Special Relativity, missing therefore the sophistication of General
Relativity. \\

A typical attempt in this regard has been suggested in Barbour J : End of Time : The Next
Revolution in Physics, Oxford University Press, 2001, where time as such is simply
eliminated completely, and all that is retained is some set of so called "now"-s. The
consequence is that, after some decades of holding to such an idea of a rather extreme
barrenness of the underlying structure, and nevertheless trying to develop it into a relevant
enough theory of Physics, not much has be achieved so far. \\
In this regard, one may note that the mentioned exclusive focus on "now" has an age old most
respected tradition among a variety of esoteric teachings across continents. However, such
teachings are not supposed to, and in fact, do not in any way aim at setting up operationally
effective theories in one's everyday practical realms, such as the theories of Physics are
expected to be. On the contrary, such teachings aim to teach one to differentiate between what
is eternal and immutable, and what on the other hand is changing. And then, for the latter,
one is simply advised to address each and every changing situation as an uninvolved
participant and according to the specifics of the need that happens to arise in the
"now" ... \\
And of course, we are all aware that such teachings have never contributed much and in a more
direct manner to the development of any science. \\

In this way the question arises :

\begin{itemize}

\item What may all the possible ways be to implement background independent and relevant
enough theories of Physics ?

\end{itemize}

A first fact to note in this regard is, perhaps, the following. \\

Going from manifestly background dependent theories, such a Newtonian Mechanics and Special
Relativity, to background independent one involves a certain {\it negation} \\
And as it happens, we humans do not seem to be particularly good with the logical operation
of negation, since it often involves a far more {\it emotional} situation than logical
operations such a "and", "or", and so on. \\
Consequently, we tend, when negating, to do it in {\it total} and {\it extreme} ways ... \\

Returning now to the issue of the negation of background dependence, and approaching the
logical operation of negation in a more careful manner, we can note the following : \\

The negation of background dependence can apparently result in at least {\it two}
alternatives : \\

1) No background at all, and thus, as so often assumed so far, no possibility of background
dependence. \\

2) No unique background, be it one universally valid as in Newtonian Mechanics and Special
Relativity, or unique to each specific mass distribution as in General Relativity. \\

And as we have seen, so far only alternative 1) has been explored, in spite of the fact that
"no background at all" can hardly be seen as anything else but yet another kind of
"background", and one which on top of it is not only unique, but also universally valid, thus
a throw back to the situation in Newtonian Mechanics and Special Relativity. \\

Therefore, we suggest the exploration of alternative 2) above, with certain corresponding
specifics mentioned in the next section. \\ \\

{\bf 9.3. Independence of Many Backgrounds as \\
          \hspace*{0.8cm} Background Independence} \\

In section 4, a more detailed consideration of what may be seen as background in theories
of Physics was presented. And it was argued that, at the present time, what may appear as more
realistic, or at least, less unrealistic in this regard, is to formulate theories of Physics
in terms of {\it scalars} given by various so called {\it reduced power algebras} which
constitute a very large family, can be constructed and used quite easily, and are natural
extensions of the usual scalars given by the field $\mathbb{R}$ of real numbers or the field
$\mathbb{C}$ of complex numbers. \\

The respective {\it simultaneous} formulation of theories of Physics in all, or in most of
such algebras of scalars would  offer the possibility of identifying which of the theories are
independent of the specific scalar algebras used, and which are not. \\

Certainly, and as mentioned, such an independence would mean a significant extension and
deepening of the Principle of Relativity. \\

However, of importance here, a further advantage would be the resulting obvious {\it
background independence} of such theories of Physics, at least as far as the large amount of
different backgrounds given by the respective scalar algebras is concerned. \\

And needless to say, reformulating theories of Physics in terms of various such scalar
algebras would be a far easier and more natural venture than building such theories from the
start within such structurally barren backgrounds as those suggested so far by various
proponents of background independent theories. Indeed, the suggested algebras of scalars are
in fact richer structures than the usual real or complex numbers, thus their use can offer the
possibility not so much of having to reinvent Physics in a structurally barren setup, but
rather to enrich it with the new possibilities available. \\ \\

{\bf Appendix} \\ \\

For the convenience of physicists, we recall here a few basic concepts on {\it filters} on
arbitrary sets, as well as from Algebra, Partially Ordered sets and convergence structures on
Algebras. The respective concepts are introduced step by step, culminating with the ones we
are mostly interested in, namely, {\it fields} and {\it algebras}, their {\it Archimedean},
respectively, {\it non-Archimedean} instances. \\
A detailed textbook to consult regarding Algebras in general is Cohn P M : Algebra, Volumes
1 and 2. Wiley, New York, 1974. \\ \\

{\bf A.1. Filters on Sets} \\

A modern and powerful concept, in spite of its intuitive simplicity,
in formulating, among others, large classes of limiting type
processes is that of {\it filter} on an arbitrary
nonvoid set, as defined next. \\

Let $\Lambda$ be a nonvoid set. A set ${\cal F}$ of subsets $I
\subseteq \Lambda$ is called
a {\it filter} on $\Lambda$, if and only if the following four conditions hold \\

$~~~~~~ {\cal F} \neq \phi $ \\

$~~~~~~\phi \notin {\cal F} $ \\

$~~~~~~ I, J \in {\cal F} ~~\Longrightarrow~~ I \cap J \in {\cal F} $ \\

$~~~~~~ I \in {\cal F},~ I \subseteq J \subseteq \Lambda
                            ~~\Longrightarrow~~ J \in {\cal F} $ \\

The meaning of the above is as follows. The subsets $I \subseteq \Lambda$ which belong to a
filter ${\cal F}$, that is, for which we have $I \in {\cal F}$, are supposed to be "large",
while their complements $\Lambda \setminus I$ are supposed to be "small" in $\Lambda$. Thus
the first above condition means that, actually, there exist "large" subsets in $\Lambda$. The
second condition means that the void subset in $\Lambda$ is not "large". The third condition
requires that the intersection of two "large" subsets is still "large". Finally, the fourth
condition requires that a set containing a "large" subset is itself "large". \\

A typical usual situation where we encounter a filter is when we define in Calculus the concept
of {\it limit} \\

$~~~~~~ \lim_{~n \to \infty}~ x_n = x $ \\

where $x_n$, with $n \in \mathbb{N}$, and $x$ are real scalars. Indeed, this definition is as
follows : \\

$~~~~~~ \forall~~ \epsilon > 0 ~:~ \exists~~ m_\epsilon \in
\mathbb{N} ~:~
        \forall~~ n \in \mathbb{N},~ n \geq m_\epsilon ~:~ |~ x - x_n ~| \leq \epsilon $ \\

This in usual intuitive terms means that $|~ x - x_n ~|$ becomes negligible for nearly all
indices $n \in \mathbb{N}$, that is for all "large" subsets $I$ of indices in $\mathbb{N}$.
Thus if we take $\Lambda = \mathbb{N}$, then its "large" subsets $I$ of indices $n$ of
interest are those for which exists a corresponding $m \in \mathbb{N}$, such that $\{~ n \in
\mathbb{N} = \Lambda ~|~ n \geq m ~\} \subseteq I$. And obviously, the set ${\cal F}$ of all
such "large" subsets $I \subseteq \Lambda = \mathbb{N}$ is a filter on $\Lambda =
\mathbb{N}$. \\

It should, however, be mentioned that filters are useful not only with respect to Calculus,
or more generally, Topology. Indeed, as is well known, they prove to be powerful tools in a
variety of branches of Mathematics. In this paper, in particular, they are used to define the
reduced power algebras, following well known ideas in Model Theory, a branch of Mathematical
Logic. \\ \\

{\bf A.2. Basic Algebraic Structures} \\

We start with an auxiliary but basic algebraic concept. Namely, a {\it group} is a structure
$( G, \alpha )$, where $G$ is a nonvoid set and \\ $\alpha : G  \times G \longrightarrow G$
is a binary operation on G which is :

\begin{itemize}

\item associative : \\

$~~~~~~ \forall~~ x, y, z \in G ~:~~ \alpha ( \alpha ( x, y ), z ) =
                                       \alpha ( x, \alpha ( y, z ) ) $ \\

\item has a neutral element $e \in G$ : \\

$~~~~~~ \forall~~ x \in G ~:~ \alpha ( x, e ) = \alpha ( e, x ) = x $ \\

\item and each element $x \in G$ has an inverse $x\,' \in G$ : \\

$~~~~~~ \alpha ( x, x\,' ) = \alpha ( x\,', x ) = e $

\end{itemize}

It is easy to see that the neutral element $e$ is unique. Also, for any given $x \in G$, its
inverse element $x\,' \in G$ is unique. \\

The group $( G, \alpha )$ is {\it commutative}, if and only if : \\

$~~~~~~ \forall~~ x, y \in G ~:~ \alpha ( x, y ) = \alpha ( y, x ) $ \\

In such a case the binary operation $\alpha$ is simply denoted by $+$ and called {\it
addition}, namely \\

$~~~~~~ \alpha ( x, y ) = x + y,~~~ x, y \in G $ \\

Further, in this commutative case, the neutral element is denoted by $0$, namely, $e = 0$,
while for every $x \in G$, its inverse $x\,'$ is denoted by $-x$. \\

It will be useful to recall the following. Given any group element $x \in G$ and any integer
number $n \geq 1$, we can define the group element $n x \in G$, by \\

$~~~~~~ n x ~=~ \begin{array}{|l}
                  ~x ~~\mbox{if}~~ n = 1 \\ \\

                  ~x + x + x + . . . + x ~~\mbox{if}~~ n \geq 2
       \end{array} $ \\

where the respective sum has $n$ terms. The meaning of this operation is easy to follow.
Namely, $n x$ can be seen as $n$ {\it steps} of {\it length} $x$ each, in the direction $x$.
This interpretation will be particularly useful in understanding the condition defining the
{\it Archimedean} property, and thus, of the {\it non-Archimedean} property as well. \\

We recall that the usual addition gives a {\it commutative group} structure on the integer
numbers $\mathbb{Z}$, as well as on the rational numbers $\mathbb{Q}$, real numbers
$\mathbb{R}$, complex numbers $\mathbb{C}$, and also on the set $\mathbb{M}^{m, n}$ of $m
\times n$ matrices of real or complex numbers, for every $m, n \geq 1$. \\

Our main interest is in the algebraic structures of {\it field} and {\it algebra}. In this
regard, we must first start with the following somewhat more general concept. A {\it ring} is
a {\it commutative group} $( S, + )$ on which a second binary operation $\beta : S \times S
\longrightarrow S$, called {\it multiplication}, is defined with the properties :

\begin{itemize}

\item $\beta $ is associative

\item $\beta$ is distributive with respect to addition : \\

$ \begin{array}{l}
           \forall~~ x, y, z \in S ~:~ \\ \\
           ~~~~~ \beta (x, y + z ) = \beta ( x, y ) + \beta ( x, z ) \\ \\
           ~~~~~ \beta ( x + y, z ) = \beta ( x, z ) + \beta ( y, z )
  \end{array} $

\end{itemize}

Usually, this second binary operation $\beta$ is called {\it multiplication}, and it
is denoted by $.$~, namely \\

$~~~~~~ \beta ( x, y ) = x . y,~~~ x, y \in S $ \\

and often, it is denoted even simpler as merely $x y = x . y$, with $x, y \in S$. \\

The ring $( S, +, . )$ is called {\it unital}, if and only if there
is an element $u \in S$,
such that \\

$~~~~~~ \forall~~ x \in S ~:~ u . x = x . u = x $ \\

Usually, the respective {\it unit} element $u \in S$ is denoted by $1$, namely \\

$~~~~~~ u = 1 $ \\

and it is easy to see that it is {\it unique}, whenever it exists. \\

The ring $( S, +, . )$ is called {\it commutative}, if and only if \\

$~~~~~~ \forall~~ x, y \in S ~:~ x . y = y . x $ \\

We recall that with the usual addition and multiplication, the
integer numbers $\mathbb{Z}$ are {\it commutative unital rings}, and
so are the rational numbers $\mathbb{Q}$, the real numbers
$\mathbb{R}$ and complex numbers $\mathbb{C}$, while the set
$\mathbb{M}^n = \mathbb{M}^{n, n}$
of $n \times n$ square matrices, with $n \geq 2$, is a {\it noncommutative} unital ring. \\

As we shall see, a crucial issue in rings, and thus in fields, and
more generally, in algebras is the possibility to perform {\it
divisions}. Indeed, as can be noted, in rings one can make arbitrary
additions, subtractions and multiplications. However, as seen
already with the $2 \times 2$ square matrices in $\mathbb{M}^2$,
division is a far more sensitive operation. In this regard, several
important concepts in rings are the following, and they can
certainly be
encountered in the case of square matrices in $\mathbb{M}^n$, with $n \geq 2$. \\

Given a ring $( S, +, . )$, an element $x \in S$ is called {\it
invertible}, or a {\it unit} - which is not to be confused with the
above concept of unit element - if and only if it has a
multiplicative inverse, that is, there exists $x\,' \in S$, such that \\

$~~~~~~ x . x\,' = x\,' . x = 1 $ \\

in which case it follows easily that $x\,'\in S$ is unique for the
respective $x \in S$.
Usually, the multiplicative inverse, when it exists, is denoted by \\

$~~~~~~ x\,' = x^{-1} $ \\

Obviously $0 \in S$ cannot have a multiplicative inverse, except in
the case when $0 = 1$, which means that the ring $S$ is trivial,
since it reduces to the single element $0$. Thus the issue is
whether or not all nonzero elements $x \neq 0$ in a given ring $S$
may have a multiplicative inverse. And in general, this is not the
case, as one can already see with the
$2 \times 2$ square matrices $\mathbb{M}^2$. For instance, the matrix \\

$~~~~~~ \left ( \begin{matrix} 1 & 0 \\ 0 & 0 \end{matrix} \right ) $ \\

is not zero, yet it has zero determinant, thus it cannot have a multiplicative inverse. \\

Now, a ring $( S, +, . )$ is called a {\it division ring}, or a {\it
skew field}, if and only
if each of its nonzero elements $x \neq 0$ has a multiplicative inverse. \\

Clearly, $\mathbb{Q}, \mathbb{R}$ and $\mathbb{C}$ are each a
division ring, while the ring $\mathbb{M}^2$ of $2\times 2$ square
matrices is {\it not} a division ring, since as we have
seen above, the nonzero matrix \\

$~~~~~~ \left ( \begin{matrix} 1 & 0 \\ 0 & 0 \end{matrix} \right ) $ \\

does {\it not} have a multiplicative inverse. \\

A second concept in rings is a follows. In a given ring $( S, +, .
)$, a nonzero element $x \neq 0$ is called a {\it zero divisor}, if
and only if there is another nonzero element $y \neq
0$, such that their product nevertheless vanishes, that is \\

$~~~~~~ x y = 0 $ \\

A nontrivial ring $( S, +, . )$ which does {\it not} have zero
divisors is called {\it entire}.
And in case the ring is also commutative, then it is called {\it integral domain}. \\

In this regard, it is easy to see that a division ring is always entire. \\

A property of importance in rings is the following {\it cancelation}
law. Given $a \in
S$ which is {\it not} a zero divisor, then  for every $x, y \in S$, we have \\

$~~~~~~ \mbox{if }~~ a x = a y ~~\mbox{or}~~ x a = y a, ~~\mbox{then}~~ x = y $ \\

The consequence of the above is that in rings with zero divisors one
cannot always simplify
factors in a product. Namely, for $x, y \in S$, the relation \\

$~~~~~ x . y = 0 $ \\

need {\it not} always imply that \\

$~~~~~ x = 0 ~~\mbox{or}~~ y = 0 $ \\

as illustrated in the sequel by the product of two matrices in
$\mathbb{M}^2$. This further
means that, given $x, y, z \in S$, the relation \\

$~~~~~~ x . y = x . z $ \\

or for that matter, the relation \\

$~~~~~~ y . x = z . x $ \\

need {\it not} always allow the simplification by $x$, thus need {\it not} always imply that \\

$~~~~~~ y = z $ \\

even if $x \neq 0$. \\

As an effect, in rings with zero divisors not every nonzero element
has an inverse. Indeed, assuming the contrary, let $x . y = 0$, with
$x, y \in S, x \neq 0$. Then there exists a multiplicative inverse
$x\,' \in S$ for $x$, which means that $x . x\,' = x\,' . x = 1$.
Hence $x\,' . ( x . y ) = x\,' . 0$, or due to the associativity of
the product, we have $( x\,' . x ) . y = 0$, which means $y = 1 . y
= ( x\,' . x ) . y = 0$. Thus we obtained that $x . y = 0$ and $x
\neq 0$ imply $y = 0$, which gives the contradiction that $S$ cannot
have zero
divisors. \\

Clearly, $\mathbb{Q}, \mathbb{R}$ and $\mathbb{C}$ are rings without
zero divisors, while the set $\mathbb{M}^2$ of $2\times 2$ matrices
has zero divisors, a fact illustrated by such a
simple example as \\

$~~~~~~ \left ( \begin{matrix} 1 & 0 \\ 0 & 0 \end{matrix} \right )~
        \left ( \begin{matrix} 0 & 0\\ 0 &  1 \end{matrix} \right ) =
        \left ( \begin{matrix} 0 & 0 \\ 0 & 0 \end{matrix} \right ) $ \\

An algebraic structure of great importance is that of {\it fields}.
A ring $(F,+, .)$ is a
field, if and only if \\

\begin{itemize}

\item every nonzero element $x \in F$ has a multiplicative inverse $x\,' \in F$, thus $F$
is a division ring, or a skew field, and

\item the multiplication in $F$ is commutative.

\end{itemize}

It follows that a field {\it cannot} have zero divisors. In this
regard, $\mathbb{Q}, \mathbb{R}$ and $\mathbb{C}$ are fields, while
$\mathbb{Z}$ and $\mathbb{M}^n$, with $n \geq 2$, are not fields.
The ring $\mathbb{Z}$ is nevertheless an integral domain. But it is
not a field, since none of its nonzero elements, except for $1$ and
$- 1$, has an inverse. On the other hand, as we have seen, the rings
$\mathbb{M}^n$, with $n \geq 2$, have zero divisors,
thus they cannot be fields. \\

Lastly, a ring $( A, +, . )$ is called an {\it algebra} over a given
field $\mathbb{K}$, if and only if there exists a third binary
operation $\gamma : \mathbb{K} \times A \longrightarrow A$, called
{\it multiplication with a scalar} in $\mathbb{K}$, namely, for each
scalar $a \in K$, and each algebra element $x \in A$, we have
$\gamma ( a, x ) \in A$. Usually, this binary operation $\gamma$ is
also written as a multiplication $.$, even if that may on occasion
cause confusion. However, one should remember that in an algebra
there are two multiplications, namely, one between two algebra
elements $x, y \in A$, and which gives the algebra element $x . y
\in A$, and another multiplication between a scalar $a \in
\mathbb{K}$
and an algebra element $x \in A$, giving the algebra element $a . x \in A$. \\

The properties of this second binary operation, namely, of
multiplication with scalars, are as follows. For $a, b \in K,~ x, y
\in A$, we have

\begin{itemize}

\item ~~~~~~ a . ( x + y ) = ( a . x ) + ( a . y )

\item ~~~~~~ ( a + b ) . x = ( a . x ) + ( b . x )

\item ~~~~~~ ( a . b ) . x = a . ( b . x )

\item ~~~~~~ 1 . x = x

\end{itemize}

Thus an {\it algebra} is in fact a structure $( A, +, ., . )$ with
one addition and two multiplications. And clearly, any field
$\mathbb{K}$ can be seen as an algebra over itself, in which case
the two multiplications are in fact the same, and not only denoted
in the same
way. \\
Otherwise, and algebra is a somewhat more general structure than a
field, and the difference
is in the stronger {\it restrictions} on division in the former. \\
Also, unlike in fields, the multiplication in algebras can be non-commutative. \\

Given a field $\mathbb{K}$, such as for instance, $\mathbb{K} =
\mathbb{R}$, or $\mathbb{K} = \mathbb{C}$, a typical and important
algebra over $\mathbb{K}$ is the set $\mathbb{M}^n_\mathbb{K}$ of $n
\times n$ square matrices with elements which are scalars in
$\mathbb{K}$, where $n \geq 2$. Here the difference between the two
multiplications in an algebra is obvious. The first multiplication
is that between two matrices $A,B \in \mathbb{M}^n_\mathbb{K}$ . The
second multiplication is that between a scalar $a \in
\mathbb{K}$ and a matrix $A \in \mathbb{M}^n_\mathbb{K}$. \\

Clearly, the multiplication between two matrices in
$\mathbb{M}^n_\mathbb{K}$ is
non-commutative, whenever $n \geq 2$. \\

At last, in order to have a better insight into the relative {\it
scarcity} of available fields, when compared with the {\it
abundance}, as well as ease to construct and use of
algebras, we recall several well known results about the former. \\

An important concept is that of the {\it characteristic} of a field
$(F,+, .)$ which is defined as follows. Let $1_F \in F$ be unit
element of the multiplication in $F$. If there is
some integer $n \geq 2$, such that \\

$~~~~~~ n . 1_F = 0 $ \\

then the field $F$ is said to have characteristic $n$, provided that
$n$ is the smallest with
that property, in which case it can be shown that $n$ must be a {\it prime} number. \\
If there is no such an integer, then $F$ is said to have characteristic zero. \\

Clearly, $\mathbb{Q}, \mathbb{R}$ and $\mathbb{C}$ are fields with
characteristic zero. On the other hand, the fields $\mathbb{Z}_p$ of
integers modulo any prime number $p \geq 2$, have
characteristic $p$. \\

The following well known result, Cohn [Vol. 1, p. 125], gives a
simple classification of all
possible fields : \\

Every field $F$ contains a smallest sub-field $F_*$ which is
isomorphic to $\mathbb{Q}$, if the characteristic of $F$ is zero,
and alternatively, it is isomorphic to $\mathbb{Z}_p$, if
the characteristic of $F$ is a prime number $p \geq 2$. \\ \\

{\bf A.3. Quotient Structures} \\

Given a commutative group $( G, + )$, there is a basic construction
which leads to certain further groups, called {\it quotient} groups
of $G$. Namely, let $H \subseteq G$ be any
subgroup of $G$. Then we define the set \\

$~~~~~~ G / H = \{~ x + H ~~|~~ x \in G ~\} $ \\

where $x + H = \{ x + y ~|~ y \in H \}$, and we note that for $x, y \in G$, we have \\

$~~~~~~ x + H = y + H ~\Longleftrightarrow~ y - x \in H $ \\

Actually, the binary relation $\approx_H$ on $G$, given for $x, y \in G$, by \\

$~~~~~~ x \approx_H y ~\Longleftrightarrow~ x + H = y + H $ \\

is an {\it equivalence} relation on $G$, that is, it is reflexive,
symmetric and transitive.
Thus in fact \\

$~~~~~~ G / H = G / \approx_H $ \\

Now the set $G / H$ can be endowed with a {\it commutative} group
structure generated by the
one given on $G$. Namely, we simply define the addition $+_H$ on $G / H$, by \\

$~~~~~~ ( x + H ) +_H ( y + H ) = ( x + y ) + H,~~~ x, y \in G $ \\

And then \\

$~~~~~~ ( G / H, +_H ) $ \\

is called the {\it quotient} group of $G$ generated by the subgroup
$H$. For simplicity, the
addition $+_H$ in the quotient group $G / H$ will be denoted by $+$.\\

What is important is that the above quotient group construction is
valid also for {\it non-commutative} groups $( G, . )$, provided
that the following mild restriction is made. Instead of an arbitrary
subgroup $H$ of $G$, we only consider {\it normal} subgroups $H$ of
$G$, namely, subgroups $H$ which satisfy the condition \\

$~~~~~~ x \,.\, H = H \,.\, x~,~~~ x \in G $ \\

The importance of the quotient group construction is, among others,
in the fact that it goes
well beyond groups. \\
Indeed, let $X$ be a ring, field or an algebra, then $X$ has a
commutative group structure with respect to the addition operation
$+$ in the respective ring, field or algebra. Therefore, if $Y
\subseteq X$ is a subgroup of $X$ in that commutative group
structure, then as above,
one can define the commutative quotient group \\

$~~~~~~ ( X / Y, + ) $ \\

The fact of interest is that, corresponding to $X$ being
respectively a ring, field or algebra, this quotient $X / Y$ will
also be a ring, field or algebra, provided that $Y$ is not only a
subgroup in $X$, but also an {\it ideal}, namely, it satisfies the condition \\

$~~~~~~ x \,.\, Y \,\cup\, Y \,.\, x \,\subseteq Y,~~~ x \in X $ \\

The ideal $Y$ in $X$ is called {\it proper}, if and only if \\

$~~~~~~ Y \subsetneqq X $ \\ \\

{\bf A.4. The Archimedean Property} \\

The Archimedean property, as much as the property of being non-
Archimedean, is essentially related to certain {\it algebraic} plus
{\it partial order} structures. A simple way to deal with the issue
is to consider ordered groups. And in fact, we can restrict
ourselves to commutative groups. Commutative groups were defined
above, therefore, here we briefly recall
the definition of {\it partial orders} and then relate them to the group structure. \\

In general, a partial order $\leq$ on an arbitrary nonvoid set $X$
is a binary relation $x \leq y$ between certain elements $x, y \in
X$, which has the following three properties

\begin{itemize}

\item it is {\it reflexive} ~:~ \\

$~~~~~~ \forall~~ x \in X ~:~ x \leq x $ \\

\item it is {\it antisymmetric} ~:~ \\

$~~~~~~ \forall~~ x, y \in X ~:~ x \leq  y,~~ y \leq x ~\Longrightarrow~ x = y $ \\

\item it is {\it transitive} ~:~ \\

$~~~~~~ \forall~~ x, y, z \in X ~:~ x \leq y,~~ y \leq z
~\Longrightarrow~ x \leq z $

\end{itemize}

In case we have the additional property \\

$~~~~~~\forall~~ x, y \in X ~:~ \mbox{either}~~ x \leq y, ~~\mbox{or}~~ y \leq x $ \\

then $\leq$ is called a {\it linear} order on $ X$. \\

Given now a commutative group $( G, + )$, a partial order $\leq$ on
$G$ is called {\it
compatible} with the group structure, if and only if \\

$~~~~~ \forall~~ x, y, z \in G ~:~ x \leq y ~\Longrightarrow~ x + z \leq y + z $ \\

A {\it partially ordered commutative group} is by definition a
commutative group $( G, + )$ together with a compatible partial
order $\leq$ on $G$. In such a case, for simplicity, we shall use
the notation $( G, +, \leq )$. In particular, we have a linearly
ordered commutative group when the compatible partial order $\leq$
is linear. It is easy to see that in the general case of a partially
ordered commutative group $( G, +, \leq )$, the above condition of
compatibility between the partial order $\leq$ and the group
structure can be simplified as
follows \\

$~~~~~~ x, y \geq 0 ~\Longrightarrow~ x + y \geq 0 $ \\

where $0 \in G $ is the neutral element in $G$. \\

We recall that $\mathbb{Z}, \mathbb{Q}$ and $\mathbb{R}$ are
commutative groups. It is now easy to see that with the usual order
relation $\leq$, each of them is a linearly ordered
commutative group. \\

Examples of partially ordered commutative groups which are {\it not}
linearly ordered are easy to come by. Indeed, let us consider the
$n$-dimensional Euclidean space $\mathbb{R}^n$, with $n \geq 2$.
With the usual addition of its vectors, this space is obviously a
commutative group. We can now define on it the partial order
relation $\leq$ as follows. Given two vectors $x = ( x_1, x_2, x_3,
. . . , x_n ), y = ( y_1, y_2, y_3, . . . , y_n ) \in \mathbb{R}^n$,
then we
define $x \leq y $ coordinate-wise, namely \\

$~~~~~~ x \leq y ~\Longleftrightarrow~  x_1 \leq y_1,~ x_2 \leq
y_2,~ x_3 \leq y_3, . . . ,
                                                  x_n \leq y_n $ \\

It is easy to see that this partial order is compatible with the
commutative group on $\mathbb{R}^n$, but it is {\it not} a linear
order, when $n \geq 2$. Indeed, this can be seen even in the
simplest case of $n = 2$, if we take $x = ( 1, 0 )$ and $y = ( 0, 1
)$, since then
we do not have either $x \leq y$, or $y \leq x$. \\

In particular, $\mathbb{C}$, as well as  $\mathbb{M}^{m,
n}_\mathbb{R}$, and $\mathbb{M}^{m, n}_\mathbb{C}$, with $m \geq 2$
or $n \geq 2$, are partially and {\it not} linearly ordered
commutative groups. Indeed, when it comes to their group structure,
each of them can be seen as an Euclidean space. Namely $\mathbb{C}$
is isomorphic with $\mathbb{R}^2$, while $\mathbb{M}^{m,
n}_\mathbb{R}$ is isomorphic with $\mathbb{R}^{m n}$, and
$\mathbb{M}^{m, n}_\mathbb{C}$ is isomorphic with $\mathbb{R}^{2 m n}$. \\

Finally, we can turn to the issue of being, or for that matter, of not being Archimedean. \\

A partially ordered commutative group $( G, +, \leq )$ is called
{\it Archimedean}, if and
only if \\

(ARCH)~~~ $ \exists~~ u \in G,~ u \geq 0 ~:~ \forall~~ x \in G,~ x
\geq 0 ~:~ \exists~~
                                                        n \in \mathbb{N} ~:~ n u \geq x $ \\

There are several alternative and not necessarily equivalent
formulations of that condition. However, the above one has the
following clear intuitive interpretation : by choosing as step
size $u \in G$, one can in a finite number $n$ of steps pass beyond any given $x \in G$. \\

Let us note that the commutative groups $\mathbb{Z}, \mathbb{Q}$ and
$\mathbb{R}$, when considered with the usual partial order $\leq$,
are each Archimedean, in the sense of (ARCH)
since one can obviously choose $u = 1$ in that condition. \\

The Archimedean property (ARCH) also holds for the commutative
groups $\mathbb{C}$ and $\mathbb{R}^n$, with $n \geq 2$, as well as
for $\mathbb{M}^{m, n}_\mathbb{R}$ and $\mathbb{M}^{m,
n}_\mathbb{C}$, with $m \geq 2$ and/or $n \geq 2$. Indeed, since
each of these commutative groups are isomorphic with a corresponding
Euclidean space, it is sufficient to show that the Euclidean spaces
$\mathbb{R}^n$, with arbitrary $n \geq 2$, are Archimedean in the
sense of (ARCH). For that, however, it is enough to note that one
can choose $u =
( 1, 1, 1, . . . , 1 ) \in \mathbb{R}^n$ in the above condition. \\

In an alternative form, instead of (ARCH), the Archimedean property is formulated as \\

(ARCH+)~~~ $ \begin{array}{l}
                        \forall~~ x \in G,~ x\geq 0 ~:~ \\ \\
                        ~~~~ \left ( \begin{array}{l}
                                            \exists~~~ y \in G ~: \\
                                            \forall~~ n \in \mathbb{N} ~: \\
                                            ~~~~ n x \leq y
                                      \end{array} \right ) ~\Longrightarrow~ x = 0
                \end{array} $ \\ \\

Here we can note that in a {\it linearly} ordered group $( G, +, \leq )$, we have \\

$~~~~~~ (ARCH+) ~~\Longrightarrow~~ (ARCH) $ \\

Indeed, assume that (ARCH) does not hold, then \\

$~~~~~~ \forall~~ x \in G,~ x \geq 0 ~:~ \exists~~ y \in G ~:~
\forall~~
                                               n \in \mathbb{N} ~:~ n x \ngeqslant y $ \\

Thus since $\leq$ is assumed to be a {\it linear} order, it follows that \\

$~~~~~~ \forall~~ x \in G,~ x \geq 0 ~:~ \exists~~ y \in G ~:~
\forall~~
                                               n \in \mathbb{N} ~:~ n x \leq y $ \\

and then (ARCH+) is obviously contradicted. \\

In the case of partially ordered groups which are not linearly
ordered, the condition (ARCH+)
is usually meant as being the Archimedean property. \\

As above, it is easy to see that the commutative groups $\mathbb{Z},
\mathbb{Q}, \mathbb{R}, \mathbb{C}$ and $\mathbb{R}^n$, with $n \geq
2$, as well as for $\mathbb{M}^{m, n}_\mathbb{R}$ and
$\mathbb{M}^{m, n}_\mathbb{C}$, with $m \geq 2$ and/or $n \geq 2$,
satisfy the condition
(ARCH+). \\

Of relevance with respect to the reduced power algebras, we can note
that {\it infinite} dimensional vector spaces, such as for instance
$\mathbb{R}^\mathbb{N}$, are {\it not} Archimedean in the sense of
(ARCH), when the natural partial order $\leq$ is considered on
these spaces. \\

Indeed, this natural partial order $\leq$ on $\mathbb{R}^\mathbb{N}$
is defined again coordinate-wise, as follows. Given $x = ( x_1, x_2,
x_3, . . . ), y = ( y_1, y_2, y_3, . . . )
\in \mathbb{R}^\mathbb{N}$, then \\

$~~~~~~ x \leq y ~\Longleftrightarrow~ x_1 \leq y_1,~ x_2 \leq y_2,~ x_3 \leq y_3, . . . $ \\

and thus it turns $\mathbb{R}^\mathbb{N}$ into a partially ordered
commutative group that,
however, is {\it not} linearly ordered, when $n \geq 2$. \\

And now, given any $u = ( u_1, u_2, u_3, . . . ) \in
\mathbb{R}^\mathbb{N}$, we can obviously
take \\

$~~~~~~ x = ( 1 . u_1, 2 . u_2, 3 . u_3, . . . ) \in \mathbb{R}^\mathbb{N} $ \\

and then clearly, the relation $x \leq n u$ will {\it not} hold for
any $n \in \mathbb{N}$,
hence condition (ARCH) is not satisfied. \\

Let us now consider a somewhat milder partial order $\precsim$ on
$\mathbb{R}^\mathbb{N}$ which is defined as follows. Given $x = (
x_1, x_2, x_3, . . . ), y = ( y_1, y_2, y_3, . . . ) \in
\mathbb{R}^\mathbb{N}$, then \\

$~~~~~~ x \precsim y ~\Longleftrightarrow~
                \left ( \begin{array}{l}
                           \exists~~ m \in \mathbb{N} ~: \\
                           ~~~~ x_m \leq y_m,~ x_{m + 1} \leq y_{m + 1},~ x_{m + 2} \leq y_{m + 2}, . . .
                       \end{array} \right ) $ \\

and with this partial order $\mathbb{R}^\mathbb{N}$ becomes again a
partially ordered
commutative group which is not linearly ordered, if $n \geq 2$. \\

Obviously, for $x, y \in \mathbb{R}^\mathbb{N}$, we have \\

$~~~~~~ x \leq y ~\Longrightarrow~  x \precsim y $ \\

As it turns out, $\mathbb{R}^\mathbb{N}$ with this partial order
$\precsim$ {\it fails} to
satisfy condition (ARCH+). \\

Indeed, let us take \\

$~~~~~~ x = ( 1, 1, 1, . . .  ),~ y = ( 1, 2, 3, . . . ) \in \mathbb{R}^\mathbb{N} $ \\

then clearly \\

$~~~~~~ \forall~~ n \in \mathbb{N} ~:~ n x \precsim y $ \\

while at the same time \\

$~~~~~~ x \geq 0,~~ x \neq 0 $ \\

Here however we can note that even {\it finite} dimensional
Euclidean spaces can turn out {\it
not} to be Archimedean, when considered with certain {\it linear} orders. \\

The simple example of the so called {\it lexicographic} order
$\dashv$ on $\mathbb{R}^2$ can already illustrate that fact. Indeed,
we recall that $\dashv$ is defined as follows. Given $x
= ( x_1, x_2 ),~ y = ( y_1, y_2 ) \in \mathbb{R}^2$, then \\

$~~~~~~ x \dashv y ~~\Longleftrightarrow~~
                             \begin{array}{|l}
                               ~~ x_1 \leq y_2 \\
                               ~~\mbox{or} \\
                               ~~ x_1 = y_1 ~~\mbox{and}~~ x_2 \leq y_2
                              \end{array} $ \\

and thus $\mathbb{R}^2$ becomes a linearly ordered commutative group. \\

Now if we take $x = ( 0, 1 ),~ y = ( 1, 0 ) \in \mathbb{R}^2$, then clearly \\

$~~~~~~ \forall~~ n \in \mathbb{N} ~:~ n x \dashv y $ \\

while at the same time \\

$~~~~~~ 0 \dashv x,~~ x \neq 0 $ \\

therefore condition (ARCH+) is {\it not} satisfied. \\

Obviously, rings, algebras and fields each have, as far as their
respective operations of addition are concerned, a commutative group
structure as part of their definition. And when a partial order is
defined to be compatible with the respective ring, algebra or field
structure, it will among other conditions be required to be
compatible with the mentioned
commutative group structure of addition. \\

Consequently, the Archimedean conditions (ARCH) or (ARCH+) on rings,
algebras and fields can be defined exclusively in terms of the
partially ordered commutative group structure of their
respective operations of addition. \\ \\

\end{document}